\documentclass[11pt]{article}
\usepackage{moriond,epsfig}

\def\Journal#1#2#3#4{{#1} {\bf #2}, #3 (#4)}

\def\ApJ{\em Astrophys. J.}
\def\ApJS{\em Astrophys. J. Supp.}
\def\AJ{\em Astronom. J.}
\def\MNRAS{\em MNRAS}
\def\PrASA{\em Proc. Astron. Soc. Aust.}
\def\NPS{\em Nat. Phys. Sci. }
\def\CS{\em Current Sci.}
\def\PIEEE{\em Proc. IEEE}
\def\IEEEAPSN{\em IEEE Ant. and Prop. Soc. Newsletter}

\def\be{\begin{equation}}
\def\ee{\end{equation}}
\def\bea{\begin{eqnarray}}
\def\eea{\end{eqnarray}}

\begin{document}
\vspace*{4cm}
\title{THE HUBBLE SPHERE HYDROGEN SURVEY}

\author{Jeffrey B. Peterson\footnote{jbp@cmu.edu} and Kevin Bandura\footnote{ace@cmu.edu} }

\address{Department of Physics, Carnegie Mellon University\\
5000 Forbes Ave,
Pittsburgh, PA 15213, USA}
\author{Ue Li Pen\footnote{pen@cita.utoronto.ca}}
\address{Canadian Institute for Theoretical Astrophysics, University of Toronto\\
60 St. George StreetToronto, Ontario, M5S 3H8, Canada}

\maketitle\abstracts{
An all sky redshift survey, using hydrogen 21 cm emission to locate galaxies, can be used to track the wavelength of baryon acoustic oscillations imprints from $z \sim 1.5$ to $z = 0$. This will allow precise determination of the evolution of dark energy. A telescope made of fixed parabolic cylindrical reflectors offers substantial benefit for such a redshift survey. Fixed cylinders can be built for low cost, and long cylinders also allow low cost fast fourier transform techniques to be used to define thousands of simultaneous beams. A survey made with fixed reflectors naturally covers all of the sky available from it's site with good uniformity, minimizing sample variance in the measurement of the acoustic peak wavelength. Such a survey will produce about a billion redshifts, nearly a thousand times the number available today. The survey will provide a three dimensional mapping of the bulk of the Hubble Sphere.
}

\section{Introduction}
If the universe today is accelerating as it expands, some anti-gravity agent, some Dark Energy, must be acting on Hubble-radius scales. On galactic scales, however, ordinary attractive gravity dominates, otherwise galaxies and galaxy clusters would never have assembled. The strength of the dark energy force must be a different function of distance than the force of ordinary gravity.  

In the past physical scales were compressed, and average matter densities were higher, so attractive gravity dominated, even on hubble-radius scales. This means the universal expansion has likely made a recent transition from deceleration to acceleration. We should be able to witness this shift by measuring the expansion rate history $H(z)$ from redshift zero to  $\sim1.5$.

A wide variety of programs have been proposed to measure the expansion history, and some of the most promising techniques involve the measurement of baryon acoustic oscillations\cite{ab}.  Before the recombination epoch, at redshift 1100, the ionized cosmic material supported acoustic oscillations. Loss of ionization at that epoch decoupled the baryonic material from the CMB photons, terminating these oscillations and establishing in the CMB the acoustic peak patterns\cite{py} \cite{eh} shown in detail by the WMAP satellite\cite{sp}.  The baryon density field was also imprinted with acoustic oscillation structure and these $\sim ~100 Mpc $ acoustic peaks have survived to today\cite{eh}. Both the 2df\cite{co} and Sloan Digital Sky Survey\cite{ei} teams report two sigma detection of acoustic peak features in the power spectrum of the low redshift galaxy density field. Since the wavelength of the acoustic oscillation spectral peak was imprinted at a particular comoving wavelength throughout the universe, it can be used as a standard ruler\cite{standardruler}. Measuring the angular and redshift-space wavelengths of the acoustic peaks over a wide range of redshifts allows measurement of the history of the expansion rate $H(z)$\cite{se03}.

Because the acoustic peak wavelength is so long, the peak can be detected in the power spectrum and measured well, even with a low-completeness survey\cite{teg98}.  The density field is sampled only where galaxies are detected, so there will be a shot noise contribution to the wavelength uncertainty\cite{gb}. There will also be an unavoidable cosmic variance uncertainty, and these two contributions are equal at about 1000 galaxies per square degree ($z=1.5$, $\Delta z = 0.2$). Assuming a Schecter mass function, this corresponds to a survey detection threshold at about M*\cite{zw}, so only the brightest one percent of the luminosity function need be detected to reach the cosmic variance limit.

The use of redshift surveys to constrain the expansion history has been studied by several authors\cite{ab,se03,gb,bg,se05}.
An all sky survey with sufficient sensitivity to detect M* galaxies at redshift 1.5 is estimated to have about 5 \% sensitivity to the equation of state of dark energy $w = p/ \rho$, and 10 \% sensitivity to
variation of $w$ across the redshift range zero to one\cite{ab,gb,bg,se05}.  The purpose of this paper is not to review these estimates, but rather to introduce a survey program that, using off the shelf technology, at modest cost, can accomplish this important measurement.

The survey telescope discussed here uses the 21 cm hyperfine emission of hydrogen, both for detection of galaxies and for measurement of their redshift. This allows the techniques and advantages of GHz-frequency radio astronomy to be applied.  Recent advances in digital signal processing technology and in mobile telephone technology, combined with proven techniques of reflector engineering make the construction of this telescope much simpler than a few years ago.

If galaxies did not evolve, and if a concordance model with $w=-1$ applies, the received 21 cm flux from a galaxy at redshift 1.5 would be about $2 \mu Jy$ (see the appendix). But galaxies do evolve, and at $z=1.5$ we are looking back to about half the current age of the universe.
In spiral galaxies gas is turned into stars over time. At the current epoch the gas mass to stellar mass ratio is about 0.1. The gas supply is heavily depleted. 
The star formation rate at $z=1.5$ is about ten times the rate today, indicating a higher gas mass fraction, but there is currently no good direct measurement of the neutral gas content at high redshift. We need a number to begin our design work, and  assume a factor three more neutral gas at redshift 1.5.

\section{Survey Telescope}

To detect the 21 cm flux from high redshift galaxies, across the entire sky, in less than a year of observation, requires a collecting area of several hundred thousand square kilometers. Such a telescope would be about a factor ten larger than any in existence today.  Fortunately, the goal of the program is an all sky survey, so inexpensive fixed cylindrical reflectors can be used.  The parameters of the telescope and survey are listed in Table 1, and the design is described below.

\begin{table}[t]
\caption{{\bf Approximate Telescope and Survey Parameters.} Presented are rough parameters, which will be refined using results of tests of prototype components, and results of simulated observations.\label{tab:telescope}}
\vspace{0.4cm}
\begin{center}
\begin{tabular}{|c|c|}
\hline
Redshift Range & 0 - 1.5\\
Frequency Range & 1500 - 500 MHz\\
Effective Area & 400,000 $m^2$\\
System Temperature & 75 K\\
Simultaneous Beams & $>$1000\\
Instantaneous Bandwidth & 200 MHz\\
Redshift Resolution & $3 x 10^{-5}$\\
Angular Resolution & 0.5 arcminute\\
Frequency Resolution & 25 Khz\\
Sky Coverage & 2 $\pi$ steradians\\
One Sigma Flux Sensitivity in one hour on target & 8 $\mu$ Jy\\

 \hline
\end{tabular}
\end{center}
\end{table}

\begin{figure}
\vskip 2.5cm
\begin{center}
\psfig{figure=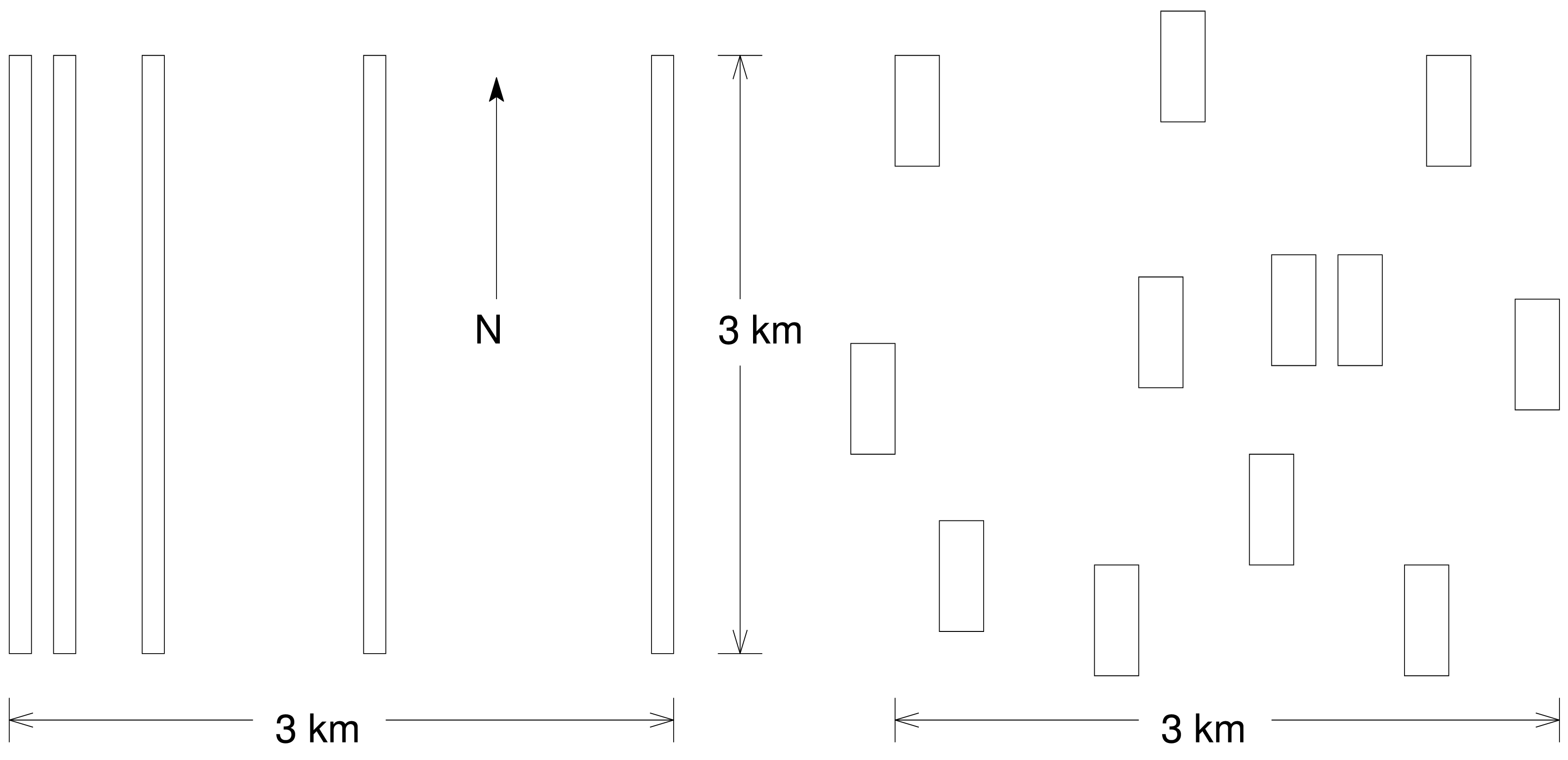,height=3.0in}
\caption{{\bf Possible Telescope Layouts} ~Left: Five long cylindrical reflectors are oriented north-south. For each cylinder a fan of thousands of simultaneous half-arc-minute (N-S) by one degree (E-W) beams is formed along the meridian.   For each beam in the fan, interferometry in the east-west direction is used to achieve half-arcminute resolution. There is some wasted reflector area at the ends of the cylinders, and this layout minimizes that waste.  Right: The same collecting area is distributed among 12 cylinders.  This layout offers lower sidelobe response and better image quality.
\label{fig:layout}}
\end{center}
\end{figure}

\subsection{Suspended Mesh Cylindrical Reflectors.}

A flexible cable tied to a support at each end, held up against gravity by its own tension, will take on a catenary shape, $\cosh  \alpha x$. This shape is very close to the parabolic shape that is needed to precisely focus radiation. The scheme discussed here uses two parallel horizontal rails as supports for square-cell welded wire mesh.  Draping the mesh from rail to rail, with the mesh grid aligned to the rails, allows individual wires of the mesh to each follow parallel catenary curves. The mesh will naturally form a cylindrical, nearly-parabolic reflector. Backside stay cables can then be added to increase the tension in the mesh so it is stable under varying wind loading. These stays can be adjusted to correct the shape to the desired parabolic cylinder.

Orienting the support rails in the North South direction allows the parabolic cylinder to focus in the East-West direction while not focusing in the North-South direction.  The line-focus of the cylinder can then be outfitted with a full-length set of small ÒfeedÓ antennas that collect the radiation from the sky. These feeds are spaced at even intervals about one half  wavelength apart. Diffraction defines the E-W angular width $\lambda/w$. In the North-South direction the reflector does not focus and the pattern is approximately that of a dipole. So, each feed sees a stripe along the Meridian, the only difference from one to the next being the North-South displacement.  If we assemble a signal that consists of the sum, with no phase difference, of all feed signals for a length $l$, we get a beam at the zenith with the N-S angular width $\sim \lambda/l$.  Using the same set of feed signals, a second beam at another North-South elevation angle can be simultaneously observed by taking a second sum, this time with a uniform phase shift from one feed to the next.  In fact, a complete fan of narrow beams, spanning the entire meridian can be simultaneously created by taking a full set of sums, or alternatively, one can fourier transform the set of feed signals versus North-South displacement. The fast fourier transform 
calculation has asymptotic order $n \log n$ rather than order $n^2$ for the summing solution, so for a long, evenly spaced set of feeds the FFT technique offers substantially lower computational cost.

\begin{figure}
\vskip 2.5cm
\begin{center}
\psfig{figure=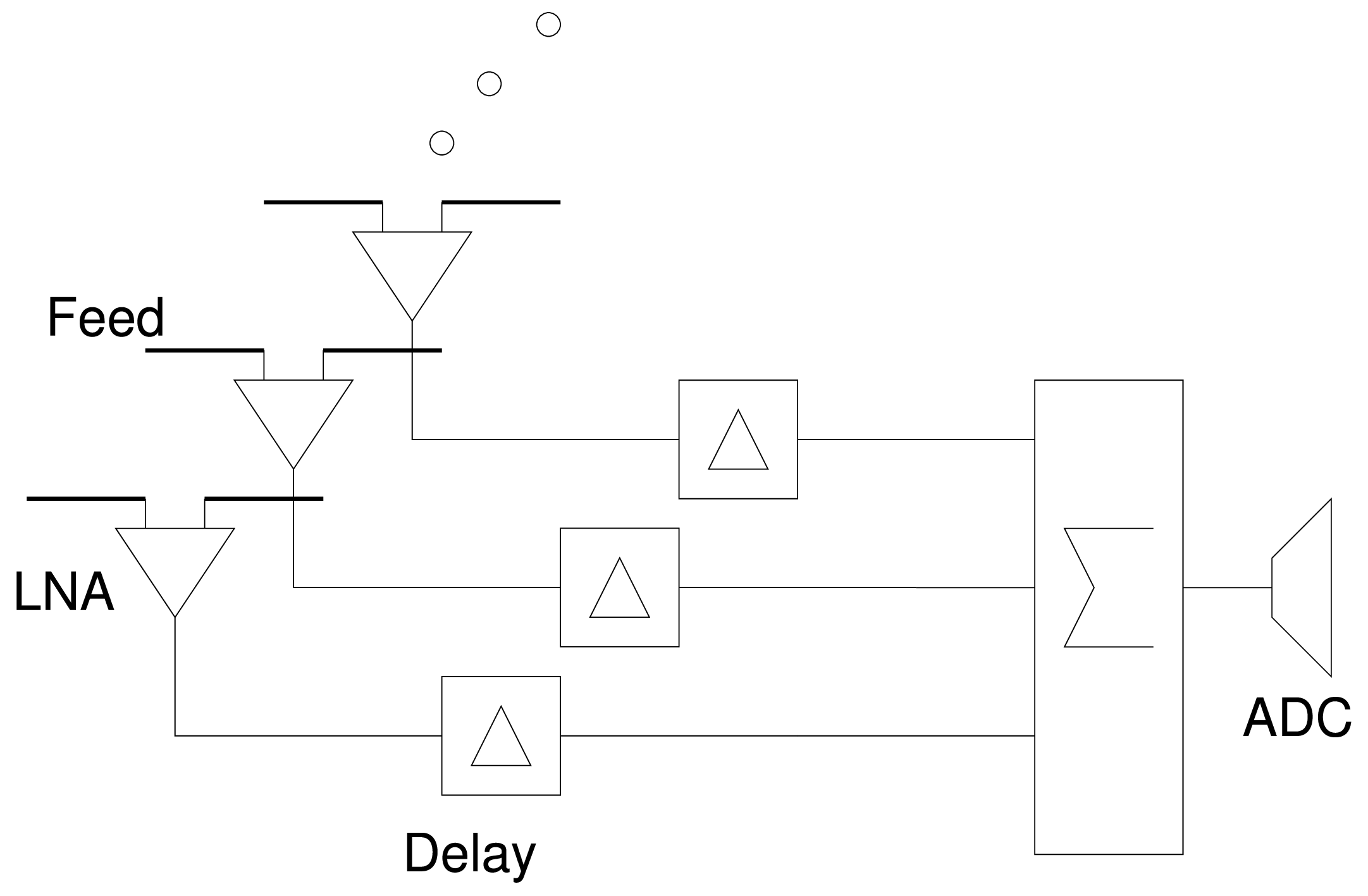,height=3.0in}
\caption{{\bf Signal summing} ~ By summing the signals from several feeds, each digitized signal can service the optimal area of cylinder, with a cylinder width of tens, rather than 100 meters.
\label{fig:layout}}
\end{center}
\end{figure}

\subsection{Large Reflector History.}

Cylindrical reflectors became popular in radio astronomy beginning in the 1950s, because they offer large collecting area at low cost. However, a cylindrical reflector needs an array of quiet amplifiers to service its line feed. In the 1980s, cryogenically cooled low noise amplifiers were developed for use in satellite downlink receivers. Using such amplifiers, reflectors curved in both dimensions (dishes) gained  the advantage over cylinders. A dish needs only one quiet amplifier, and the quiet amplifiers were expensive.  Now, low noise HEMT amplifiers, which operate at room temperature, are available for a few dollars each, and cylindrical layouts are again economical. At the same time, computing speeds have advanced to the point that inexpensive PC computers can carry out FTT operations real-time, with bandwidths of hundreds of Megahertz. It is now reasonable to use a cylindrical reflector to observe over 1000 simultaneous beams.

Mesh reflectors are common on radio telescopes for the GHz frequency range.  Mesh with spacing ~1 cm, when used at wavelengths longer than 20 cm, acts as a high efficiency reflector, one which allows wind to pass through. Also, infrared and optical sunlight are not significantly focused by the mesh, preventing solar-radiation damage to the feed system.  GMRT\cite{sw91}, Ooty\cite{sw71}, Illinois\cite{swenson}, and Molonglo\cite{mills} all use mesh reflectors.

The Arecibo telescope uses multiple cables to support the zenith-pointed spherical dish\cite{kildal}. The feed system moves to track objects. Objects can be tracked for a few hours, but must pass close to the zenith to be observed at all.  North-South cylinders allow the entire sky to be observed, with no moving parts, however any particular object is available for only a short time each day. 

\subsection{Survey Speed.}

The flux sensitivity of a radio telescope is $\Delta S \sim  2 k T_{sys}/ A \sqrt{ 2 \Delta t \Delta \nu}$, where $k$ is Boltzmann's constant, $T_{sys}$ is the noise temperature of the system, $A$ is the effective area of the telescope, $t$ is the observing time, and $\Delta \nu$ is the frequency width of the line being observed. The survey will cover solid angle $\Omega _{survey}$, repeating this coverage each day. However, at any instant the telescope sees only a strip along the meridian of solid angle $\Omega_{tel}$. The survey speed, the reciprocal of the time needed  to reach the required flux limit is $SP \sim [ \frac {\Delta S} { 2 kT_{sys}}]^2 [\frac{\Delta \nu} { \Omega_{survey}}]A^2 \Omega_{tel} $.  The factors in brackets are not easily adjustable, these values are set by the desired result and the amplifier noise temperature.  The last two factors can be chosen for engineering or financial reasons. The telescope soild angle $\Omega_{tel}$, also known as the instantaneous field of view, is proportional to the number of signals being processed simultaneously. 

Cost optimization is straightforward for fixed cylinder telescopes, since the cylinder width can be adjusted. Survey speed is maximized when twice as much is spent on collecting area (steel) as is spent on signal processing (silicon). Assuming cost coefficients \$20 per square meter of collecting area and \$500 per signal processor channel, each processed signal needs 50 square meters of collecting area.  If we digitize and process the signal from every feed, with feeds spaced 30 cm, the cylinder width would be 166 m. If it is found that maintaining the required surface precision ( $<$ 1 cm) across such a large span entails substantial additional cost, there is another alternative. One can use a stage of analog summing, shown schematically in Figure 2, before digitization. The width of the cylinder can then be reduced by increasing the number of feed signals summed before digitizing. Such a system is used on  narrow cylinders like Ooty and Molonglo. Note that the cost to process a signal will come down substantially over the next few years, eventually eliminating the need for analog summing.

\subsection{No Galaxy Detection Threshold}

Rather than use a galaxy catalog, we will calculate sky-structure power spectra directly from the measured 21 cm flux. The use of a catalog, restricting the data set to high significance detections, throws away useful signal. Because neutral hydrogen is concentrated in galaxy-size clumps, it makes sense to pass the signal stream through filters that emphasize the spatial and spectral pattern of a galaxy, but the galaxies need not be cataloged. 
Our no-threshold approach is roughly equivalent to including in a catalog all galaxies with signal to noise greater than one,
rather than the usual practice of including only 3 or 5 sigma detections. While we will not use a catalog to detect the acoustic peaks, we will still produce a catalog for use in other analyses. 

\section{Conclusions}

Fixed cylindrical reflector technology can be used to build a low cost radio telescope capable of measuring redshifts of galaxies across the sky out to redshift 1.5. The use of FFT beam forming, which is a natural choice when using cylindrical
reflectors, further reduces cost. Data from this survey can be used to detect and measure baryon acoustic oscillations, sharply constraining dark energy models.

\section*{Acknowledgments}
This work was supported in the US by grant AST 0507665 from the NSF, and in Canada by funds from NSERC.

\section*{Appendix: Sensitivity Calculations}

The emissivity of hydrogen is

\begin{equation}
\epsilon_{\nu} = \frac{1}{4 \pi} h \nu_{12} A_{12} \frac{N_2}{N_H} N_H \phi(\nu)
\end{equation}

where the transition frequency $\nu_{12} = 1420$ MHz, the Einstein coefficient $A_{12} = 2.85 \times 10^{-15} \textrm{s}^{-1}$, $N_H$ is the total number of hydrogen atoms in the galaxy, $N_2$ is the number of hydrogen atoms in the excited state, and $\phi(\nu)$ is the line profile, which we treat as a delta function.

Since $h\nu_{12}/k = 0.06 K$, even the CMB temperature is much higher than the transition energy, and the system is in the high temperature regime. The upper state is a triplet and the lower a singlet, so $N_2/N_H \approx 3/4$.  

The monochromatic luminosity is
\begin{equation}
L_{\nu} = \int_{\Omega} \epsilon_{\nu} d\Omega = \frac{3}{4} h \nu A_{12} \frac{M_{HI}}{m_H} \phi(\nu)
\end{equation}
where ${M_{HI}}$ is the mass of neutral hydrogen, and ${m_H} $ is tha mass of the hydrogen atom.
The monochromatic flux density is related to the monochromatic luminosity by
\begin{equation}
S_{\nu} = L_{\nu(1+z)}\frac{1}{4 \pi (1+z)^3 D_A^2(z)}
\end{equation}
Assuming a flat universe, the angular diameter distance   
\begin{equation}
D_A(z) = \frac{c}{1+z} \int_{0}^{z} \frac{dz}{H(z)}
\end{equation}
and
\begin{equation}
H(z) = H_o \sqrt{\Omega_m \left(1+z\right)^3 + \Omega_{\Lambda} e^{\left[3 \int_{0}^{z} \frac{1 + w(z)}{1+z} dz\right]}  }
\end{equation}
Integrating both sides of equation 3 gives
\begin{equation}
\int S_{\nu} d\nu = \frac{3}{4} h  A_{12} \frac{M_{HI}}{m_H} \frac{1}{4 \pi (1+z)^3 D_A^2(z)} \int \nu \phi(\nu) d\nu
\end{equation}
and the received flux in the 21 cm line is
\begin{equation}
S_{obs} \Delta \nu = \frac{3}{4} h \nu_{12} A_{12} \frac{M_{HI}}{m_H} \frac{1}{4 \pi (1+z)^4 D_A^2(z)} 
\end{equation}

Using the low-redshift luminosity function of Zwann etal \cite{zw},  $M_{HI} = M_{*} = 1.23 \times 10^{40} kg$ . Placing this galaxy at redshift 1.5, and adopting $\Delta \nu = 0.568 \textrm{MHz}$, $\Omega_m = 0.3$, $\Omega_{\Lambda}=0.7$ and $w=-1$ the observed flux would be $2.09 \mu\textrm{Jy}$ .

The flux limit of a radio astronomy observation is
\begin{equation}
S_{lim} = \frac{2 k T_{sys}}{A_{eff} \sqrt{2 \Delta \nu t}}
\end{equation}
where t is the integration time, $T_{sys} = 75 K$ is the system temperature and $A_{eff} = 400,000 m^2$ is the effective area. 

Using these values, and assuming that galaxies were three times more luminous at  redshift 1.5 than they are today, the required on-target integration time for signal to noise ratio one is about 6000 seconds.  If the telescope sees a strip along the meridian of width 0.2 degree, it would take $\approx 130$ days to detect all $M_{*}$ galaxies on the sky at redshift 1.5.

\end{document}